\documentclass{iaus}
\usepackage{graphicx}

\newcommand{\apj}       {\emph{ApJ}}

\newcommand{\smyr}      {{ M_\odot\ \rm yr^{-1}}}
\newcommand{\sm}        {{ M_\odot}}

\newcommand{\beq}       {\begin{equation}}
\newcommand{\eeq}       {\end{equation}}
\newcommand{\beqa}      {\begin{eqnarray}}
\newcommand{\eeqa}      {\end{eqnarray}}

\newcommand{\fkep}      {f_{\rm Kep}}

\newcommand{\msd}       {m_{*d}}
\newcommand{\msdo}       {m_{*d,\, 0}}
\newcommand{\mdsd}      {\dot m_{*d}}
\newcommand{\mdsdo}      {\dot m_{*d,\, 0}}

\newcommand{\esd}       {\epsilon_{*d}}
\newcommand{\esdb}       {\bar\epsilon_{*d}}

\newcommand{\mst}       {m_{*,\,2}}
\newcommand{\msdt}      {m_{*d,\,2}}

\newcommand{\teff}      {T_{\rm eff}}

\def\ion#1#2{#1$\;${\small\rm II}\relax}
\def\lesssim{\mathrel{\hbox{\rlap{\hbox{\lower4pt\hbox{$\sim$}}}\hbox{$<$}}}}
\def\gtrsim{\mathrel{\hbox{\rlap{\hbox{\lower4pt\hbox{$\sim$}}}\hbox{$>$}}}}

\title[Population III.1 stars] 
{Population III.1 stars: formation, feedback and evolution of the IMF}

\author[Jonathan C. Tan]   
{Jonathan C. Tan$^1$
}

\affiliation{$^1$Dept. of Astronomy, University of Florida, Gainesville, FL 32611, USA\\ email: {\tt jt@astro.ufl.edu}}

\pubyear{2008}
\volume{255}  
\pagerange{119--126}
\jname{Low-Metallicity Star Formation: From the First Stars to Dwarf Galaxies}
\editors{L.K. Hunt, S. Madden \& R. Schneider, eds.}
\begin{document}

\maketitle

\begin{abstract}
I discuss current theoretical expectations of how primordial, Pop
III.1 stars form. Lack of direct observational constraints makes
this a challenging task. In particular predicting the mass of
these stars requires solving a series of problems, which all affect,
perhaps drastically, the final outcome. While there is general
agreement on the initial conditions, $\rm H_2$-cooled gas at the
center of dark matter minihalos, the subsequent evolution is more
uncertain. In particular, I describe the potential effects of dark
matter annihilation heating, fragmentation within the minihalo,
magnetic field amplification, and protostellar ionizing
feedback. After these considerations, one expects that the first stars
are massive $\gtrsim 100\sm$, with dark matter annihilation heating
having the potential to raise this scale by large factors. Higher
accretion rates in later-forming minihalos may cause the
Pop III.1 initial mass function to evolve to higher masses.
\keywords{stars: formation, galaxies: formation, dark matter, cosmology: theory}
\end{abstract}

\firstsection 
\section{Introduction: The Importance of Pop III.1 Stars and their IMF}

The first, essentially metal-free (i.e. Population III), stars are
expected to have played a crucial role in bringing the universe out of
the dark ages: initiating the reionization process, including the
local effects of their \ion{H}{2} regions in generating shocks and
promoting formation of molecular coolants in the relic phase;
photodissociating molecules; amplifying magnetic fields to possibly
dynamically important strengths; and generating the mechanical
feedback, heavy elements and possible neutron star or black hole
remnants associated with supernovae. In these ways Pop III stars laid
the foundations for galaxy formation, including supermassive black
holes and globular clusters. Many of these processes are theorized to
depend sensitively on the initial mass function (IMF) of Pop III
stars, thus motivating its study. The formation of the first Pop III
stars in a given region of the universe is expected to have been
unaffected by other astrophysical sources and these have been termed
Pop III.1, in contrast to Pop III.2 (McKee \& Tan 2008, hereafter
MT08). Pop III.1 are important for influencing the initial conditions
for future structure formation and for having their properties
determined solely by cosmology. There is also the possibility,
described in Sect.~\ref{S:DM}, that Pop III.1 star formation may be
sensitive to the properties of weakly interacting massive particle
(WIMP) dark matter.

Unfortunately, at the present time and in the near future we expect
only indirect observational constraints on the Pop III IMF. The epoch
of reionization can be constrained by CMB polarization (Page et
al. 2007) and future high redshift 21~cm HI observations (e.g. Morales
\& Hewitt 2004). Metals from individual Pop III supernovae may have
imprinted their abundance patterns in very low metallicity Galactic
halo stars (Beers \& Christlieb 2005) or in the Ly-$\alpha$ forest
(Schaye et al. 2003; Norman, O'Shea, \& Pascos 2004). Light from the
first stars may contribute to the observed NIR background intensity,
(e.g. Santos, Bromm, \& Kamionkowski 2002), and its fluctuations
(Kashlinsky et al. 2004; c.f. Thompson et al. 2007). If massive,
supernovae marking the deaths of the first stars may be observable by
JWST (Weinmann \& Lilly 2005). If these supernovae produce gamma-ray
bursts then these may already be making a contribution to the
population observed by SWIFT (Bromm \& Loeb 2002).


The lack of direct observations of Pop III star formation means
theoretical models lack constraints, which is a major problem for
treating such a complicated, nonlinear process. Numerical simulations
have been able to start with cosmological initial conditions and
advance to the point of protostar formation (see Yoshida et al., these
proceedings), but progressing further through the protostellar
accretion phase requires additional modeling of complicated
processes, including a possible need to include extra physics such as
WIMP annihilation and magnetic fields. Building up a prediction of the
final mass achieved by the protostar, i.e. the initial mass (function)
of the star (population), is akin to building a house of cards: the
reliability of the structure becomes more and more precarious.

In this article we summarize theoretical attempts to understand the
formation process and resulting IMF of Pop III.1 stars. We have
reviewed much of these topics previously (Tan \& McKee 2008), so here
we concentrate on a discussion of some of the more uncertain aspects
in these models, including the potential effects of WIMP
annihilation on Pop III.1 star formation, fragmentation during Pop
III.1 star formation, the generation of magnetic fields, the
uncertainties in predicting the IMF from feedback models, and the
evolution of the Pop III.1 IMF. Note, when discussing possible
fragmentation during the formation of a Pop III.1 star, we will
consider all stars that result from the same minihalo to be Pop III.1,
i.e. they are unaffected by astrophysical sources external to their own
minihalo.



\section{Initial Conditions and Possible Effects of WIMP Annihilation}

The initial conditions for the formation of the first stars are
thought to be relatively well understood: they are determined by the
growth of small-scale gravitational instabilities from cosmological
fluctuations in a cold dark matter universe. The first stars are
expected to form at redshifts $z\sim 10-50$ in dark matter
``minihalos'' of mass $\sim 10^6 M_\odot$ (Tegmark et al. 1997).  In
the absence of any elements heavier than helium (other than trace
amounts of lithium) the chemistry and thermodynamics of the gas are
very simple. Once gas collects in the relatively shallow potential
wells of the minihalos, cooling is quite weak and is dominated by the
ro-vibrational transitions of trace amounts of $\rm H_2$ molecules
that cool the gas to $\sim 200$~K at densities $n_{\rm H} \sim
10^4\:{\rm cm^{-3}}$ (Abel, Bryan, \& Norman 2002; Bromm, Coppi, \&
Larson 2002). Glover et al. (these proceedings) review the effects of
other potential coolants, finding they are small for Pop III.1 star
formation.

As the gas core contracts to greater densities, the $\rm H_2$ cooling
becomes relatively inefficient and the temperature rises to $\sim
1000$~K. At densities $\sim 10^{10}\:{\rm cm^{-3}}$ rapid 3-body
formation of $\rm H_2$ occurs, creating a fully molecular region that
can cool much more efficiently. This region starts to collapse
supersonically until conditions become optically thick to the line and
continuum cooling radiation, which occurs at densities $\sim
10^{17}\:{\rm cm^{-3}}$.  Recent 3D numerical simulations have
advanced to densities of order $10^{21}$~cm$^{-3}$ (see contribution
by Yoshida et al., these proceedings), but have trouble proceeding
further given the short timesteps required to resolve the dynamics of
the high density gas of the protostar. Further numerical progress can
be achieved by introducing sink particles (Bromm \& Loeb 2004) or with
1D simulations (Omukai \& Nishi 1998; Ripamonti et al. 2002).

Alternatively, given the above initial conditions, the subsequent
accretion rate to the protostar can be calculated analytically (Tan \&
McKee 2004, hereafter TM04).  The accretion rate depends on the
density structure and infall velocity of the gas core at the point
when the star starts to form. Omukai \& Nishi (1998) and Ripamonti et
al. (2002) showed that the accreting gas is isentropic with an
adiabatic index $\gamma\simeq 1.1$ due to H$_2$ cooling; i.e., each
mass element satisfies the relation $P=K\rho^\gamma$ with the
``entropy parameter'' $K=$~const.  In hydrostatic equilibrium---and
therefore in a subsonic contraction---such a gas has a density profile
$\rho \propto r^{-k_\rho}$ with $k_\rho\simeq 2.2$, as is seen in
simulations.
TM04 describe the normalization of the core density
structure via the ``entropy parameter''
\beq 
K'\equiv (P/\rho^\gamma)/1.88\times 10^{12}~{\rm cgs}=
(\teff'/300~{\rm K})(n_{\rm H}/10^4{\rm cm}^{-3})^{-0.1},
\label{eq:kp}
\eeq
where $\teff' \equiv T+\mu\sigma^2_{\rm turb}/k$ is an effective
temperature that includes the modest effect of subsonic turbulent motions
that are seen in numerical simulations (Abel et al. 2002).

For the infall velocity at the time of protostar formation,
simulations show the gas is inflowing subsonically at about a third of
the sound speed (Abel et al 2002). Hunter's (1977) solution for mildly
subsonic inflow (Mach number =0.295) is the most relevant for this
case. It has a density that is 1.189 times greater than a singular
isothermal sphere (Shu 1977) at $t=0$, and an accretion rate that is
2.6 times greater.

Feedback from the star, whether due to winds, photoionization, or
radiation pressure, can reduce the accretion rate of the star. TM04
and MT08 define a hypothetical star$+$disk mass, $\msdo$, and
accretion rate, $\mdsdo$, in the absence of feedback. In this case,
the star$+$disk mass equals the mass of the part of the core (out to
some radius, $r$, that has undergone inside-out collapse) from which
it was formed, $\msdo=M(r)$.  The instantaneous and mean star
formation efficiencies are $\esd\equiv \mdsd / \mdsdo$ and $\esdb
\equiv \msd / \msdo = \msd/M$, respectively.

Assuming the Hunter solution applies for a singular polytropic sphere
with $\gamma=1.1$, the accretion rate is then (TM04)
\beq
\label{eq:mdot}
\mdsd=0.026 \epsilon_{*d}K'^{15/7} (M/\sm)^{-3/7}~M_\odot~{\rm yr}^{-1},
\eeq
with the stellar mass smaller than the initial enclosed core mass via
$m_* \equiv \msd/(1+f_d) =\esdb M/(1+f_d)$. We choose a fiducial value
of $f_d=1/3$ appropriate for disk masses limited by enhanced viscosity
due to self-gravity.

\subsection{Possible Effects of Dark Matter Annihilation}\label{S:DM}


Pop III.1 stars form at the centers of dark matter (DM)
minihalos. While the mass density is dominated by baryons inside $\sim
1$~pc, adiabatic contraction ensures that there will still be a
peak of DM density co-located with the baryonic
protostar. As discussed by Spolyar et al. (2008), if the dark matter
consists of a weakly interacting massive particle (WIMP) that self
annihilates, then this could lead to extra heating that can help
support the protostar against collapse. Spolyar et al. calculated
that, depending on the dark matter density profile, WIMP mass, and
annihilation cross section, the local heating rate due to dark matter
could exceed the baryonic cooling rate for densities $n_{\rm H}\gtrsim
10^{14}\:{\rm cm^{-3}}$, corresponding to scales of about 20~AU from
the center of the halo/protostar.

Natarajan, Tan, \& O'Shea (2008) revisited this question by
considering several minihalos formed in numerical simulations. While
there was some evidence for adiabatic contraction leading to a
steepening of the dark matter density profiles in the centers of the
minihalos, this was not well resolved on the scales where heating may
become important. Thus various power law ($\rho_\chi\propto
r^{-\alpha_\chi}$) extrapolations were considered for the DM
density. A value of $\alpha_\chi\simeq1.5$ was derived based on the
numerically well-resolved regions at $r\sim1$~pc. A steeper value of
$\alpha_\chi\simeq 2.0$ was derived based on the inner regions of the
simulations. In the limit of very efficient adiabatic contraction, one
expects the dark matter density profile to approach that of the
baryons, which would yield $\alpha_\chi\simeq2.2$.
For the density profiles with $\alpha_\chi\simeq 2.0$, Natarajan et
al. (2008) found that dark matter heating inevitably becomes
dominant. Natarajan et al. also considered the global
quasi-equilibrium structures for which the total luminosity generated
by WIMP annihilation that is trapped in the protostar, $L_{\chi,0}$, equals that
radiated away by the baryons, assuming both density distributions are
power laws truncated at some radius, $r_c$, with a constant density
core. This core radius was varied to obtain the equilibrium
luminosity. Typical results were $L_{\chi,0}\sim
10^3\:L_\odot$ and $r_c\simeq$ to a few to a few tens of AU.

These scales at which equilibrium is established are important for
determining the subsequent evolution of the protostar, which will
continue to gain baryons and probably additional dark matter via
adiabatic contraction. Even in the limit where no further dark matter
becomes concentrated in the protostar, that which is initially present
can be enough to have a major influence on the subsequent protostellar
evolution. As the protostar gains baryonic mass it requires a greater
luminosity for its support. If there was no dark matter heating, the
protostar would begin to contract once it becomes older than its local
Kelvin-Helmholz time, i.e. on timescales much longer than the stellar
dynamical time. If dark matter is present, it will become concentrated
as the protostar contracts, and the resulting annihilation luminosity
will grow as $L_\chi \simeq L_{\chi,0} (r_*/r_{*,0})^{-3}$, assuming a
homologous density profile. For a starting luminosity of $L_{\chi,0} =
1000\:L_\odot$ and radius of $r_{*,0}\simeq r_c = 10$~AU, this can
mean luminosities that are easily large enough to support
$\sim100\:M_\odot$ stars, i.e. $\sim10^6\:L_\odot$, at sizes of $\sim
1$~AU, i.e. much greater than their main sequence radii, which would
be $\simeq 5 R_\odot =0.02$~AU. These estimates are of course very
sensitive to the initial size of the protostar.

Full treatment of the protostellar evolution (see Freese et al. 2008
for an initial model) requires a model for the evolution of the
stellar DM content, which grows by accumulation of surrounding
WIMPs, but also suffers depletion due to the annihilation process. The
mean depletion time in the star is $t_{\rm dep} =
(\rho_\chi/\dot{\rho}_\chi) \simeq m_\chi/(\rho_\chi <\sigma_a v>)
\rightarrow 105(m_\chi/100~{\rm GeV}) (\rho_\chi/10^{12}{\rm
GeV\:cm^{-3}})^{-1}\:{\rm Myr}$, where we have normalized to typical
values of $\rho_\chi$ in the initial DM core (Natarajan et
al. 2008). If the protostar contracts from an initial radius of 10~AU
to 1~AU then $t_{\rm dep}\simeq 10^5\:{\rm yr}$. This becomes
comparable to the growth time of the protostar (i.e. the time since
its formation, its age), $t_*=2.92 \times 10^4
K'^{-15/7}(m_*/100M_\odot)^{10/7} \: {\rm yr}$ (TM04).
We see that, if replenishment of WIMPs in the protostar is negligible,
then depletion can become important for AU scale protostars of $\sim
100\:M_\odot$.

Protostars swollen by DM heating would have much cooler photospheres
and thus smaller ionizing feedback than if they had followed standard
protostellar evolution leading to contraction to the main sequence by
about 100~$M_\odot$. Ionizing feedback is thought to be important in
terminating accretion and thus setting the Pop III.1 IMF (MT08; see
\S\ref{S:feedback} below). The reduced ionizing feedback of DM-powered
protostars may allow them to continue to accrete to much higher masses
than would otherwise have been achieved.

\section{Protostellar Accretion and Disk Fragmentation}\label{S:accfrag}

Another process that may affect the IMF of the first stars is
fragmentation of the infalling gas after the first protostar has
formed. TM04 and MT08 considered the growth and evolution of the
protostar in the case of no fragmentation (and no DM heating): the
final mass achieved by the protostar is expected to be $\sim
100-200\sm$ and set by a balance between its ionizing feedback and its
accretion rate through its disk (\S\ref{S:feedback}).

The accretion disk of the protostar 
does present an environment in which density fluctuations can grow,
since there will typically be many local dynamical timescales before
the gas is accreted to the star. TM04 calculated the expected disk size, $r_d(m_*)$,
assuming conservation of angular momentum inside the sonic point,
$r_{\rm sp}$, of the inflow, finding \beq r_d =
1280\left(\frac{\fkep}{0.5}\right)^2\left(
\frac{\msdt}{\esdb}\right)^{9/7}K'^{-10/7}~{\rm AU} \rightarrow
1850\left(\frac{\fkep}{0.5}\right)^2 \frac{\mst^{9/7}}{K'^{10/7}}~{\rm
AU}
\label{eq:rd}
\eeq where $m_{*d,2}=m_{*d}/100\sm$, $m_{*,2}=m_*/100\sm$, the
$\rightarrow$ is for the case with $f_d=1/3$ and $\fkep\equiv v_{\rm
rot}(r_{\rm sp}) / v_{\rm Kep}(r_{\rm sp})$, with a typical value of
0.5 seen in numerical simulations.

The high accretion rates of primordial protostars make it likely that
the disk will build itself up to a mass that is significant compared
to the stellar mass. At this point the disk becomes susceptible to
global ($m=1$ mode) gravitational instabilities (Adams, Ruden, \& Shu
1989; Shu et al. 1990), which are expected to be efficient at driving
inflow to the star, thus regulating the disk mass. Thus TM04 assumed a
fixed ratio of disk to stellar mass, $f_d=1/3$.

Accretion through the disk may also be driven by local instabilities,
the effects of which can be approximated by simple Shakura-Sunyaev
$\alpha_{\rm ss}$-disk models. Two dimensional simulations of clumpy,
self-gravitating disks show self-regulation with $\alpha_{\rm
ss}\simeq (\Omega t_{\rm th})^{-1}$ up to a maximum value $\alpha_{\rm
ss}\simeq 0.3$ (Gammie 2001), where $\Omega$ is the orbital angular
velocity, $t_{\rm th}\equiv \Sigma kT_{\rm c,d}/(\sigma T_{\rm
eff,d}^4)$ is the thermal timescale, $\Sigma$ is the surface density,
$T_{\rm c,d}$ is the disk's central (midplane) temperature, and
$T_{\rm eff,d}$ the effective photospheric temperature at the disk's
surface. 

Gammie (2001) found that fragmentation occurs when $\Omega t_{\rm
th}\lesssim 3$. This condition has the best chance of being satisfied
in the outermost parts of the disk that are still optically
thick. However, Tan \& Blackman (2004, hereafter TB04) considered the
gravitational stability of constant $\alpha_{\rm ss}=0.3$ disks fed at
accretion rates given by eq. \ref{eq:mdot} and found that the
optically thick parts of the disk remained Toomre stable ($Q>1$)
during all stages of the growth of the protostar. Note that the
cooling due to dissociation of $\rm H_2$ and ionization of H was
included in these disk models.

We therefore expect that during the early stages of typical Pop III.1
star formation, the accretion disk will grow in mass and mass surface
density to a point at which gravitational instabilities, both global
and local, act to mediate accretion to the star. The accretion rates
that can be maintained by these mechanisms are larger than the infall
rates of eq.~\ref{eq:mdot}, and so the disk does not fragment. 

We note that if fragmentation does occur and leads to formation of
relatively low-mass secondary protostars in the disk, then one
possible outcome is the migration of these objects in the disk until
they eventually merge with the primary protostar. The end result of
such a scenario would not be significantly different from the case of
no fragmentation. Another possibility is that a secondary fragment
grows preferentially from the circumbinary disk leading to the
formation of a massive twin binary system (Krumholz \& Thompson
2007). If both stars are massive, this star formation scenario would
be qualitatively similar to the single star case in terms of the
effect of radiative feedback limiting accretion. A massive binary
system would mean that the accreting gas needs to lose less angular
momentum and binary-excited spiral density waves provide an
additional, efficient means to transfer angular momentum, compared to
the single star case. For close binaries, new stellar evolution
channels would be available involving mass transfer and merger, with
possible implications for the production of rapidly rotating
pre-supernova progenitors and thus perhaps gamma-ray bursts.




Fragmentation will only be significant for the IMF if it occurs
vigorously and leads to a cluster of lower mass stars instead of a
massive single or binary system. Clark, Glover, \& Klessen (2008)
claimed such an outcome from the results of their smooth particle
hydrodynamical simulation of the collapse of a primordial
minihalo. They allowed dense, gravitationally unstable gas to be
replaced by sink particles. They found a cluster of 20 or so
protostars formed.  As discussed by Clark et al. (2008) (see also
Glover et al., these proceedings), there are a number of caveats
associated with this result. The initial conditions (a sphere of
radius 0.17~pc with an uniform particle density of $5\times 10^5\:{\rm
cm^{-3}}$, and ratios of rotational and turbulent energy to
gravitational of 2\% and 10\%, respectively) were not derived from ab
initio simulations of cosmological structure formation. In particular,
cosmologically-formed minihalos evolve towards structures that have very
steep density gradients, centered about a single density peak. This is
likely to allow the first, central protostar to initiate its formation
long before other fluctuations have a chance to develop. The
development of a massive central object will create tidal forces in
the surrounding gas that will make it more difficult for gravitational
instabilities to develop. Furthermore, the surrounding gas is
infalling on about a local free fall time, so density perturbations
have few local dynamical timescales in which to grow. Another caveat
with the Clark et al. fragmentation results is the use of a simple
tabulated equation of state, in which gas can respond instantaneously
to impulses that induce cooling. This, and the form of the equation of
state used, lead to near isothermal conditions in the fragmenting
region.


\section{Magnetic Fields and Hydromagnetic Outflows}\label{S:magnetic}

TB04 considered the growth of magnetic fields in the accretion disk of
Pop III.1 protostars. They estimated minimum seed field strengths
$\sim 10^{-16}G$. Xu et al. (2008) have recently reported field
strengths of up to $10^{-9}G$ generated by the Biermann battery
mechanism in their simulations of minihalo formation. Such seed fields
are expected to be amplified by turbulence in the disk, attaining
equipartition strengths by the time the protostar has a mass of a few
solar masses or so. If the turbulence generates large scale helicity,
as in the model of Blackman \& Field (2002), then this can lead to the
creation of dynamically-strong fields that are ordered on scales large
compared to the disk. Such fields, coupled to the rotating accretion
disk, are expected to drive hydromagnetic outflows, such as disk winds
(Blandford \& Payne 1982).

TB04 then considered the effect of such outflows on the accretion of
gas from the minihalo, following the analysis of Matzner \& McKee
(2000). The force distribution of centrifugally-launched hydromagnetic
outflows is collimated along the rotation axes, but includes a
significant wider-angled component. Using the sector approximation,
TB04 found the angle from the rotation/outflow axis at which the
outflow had enough force to eject the infalling minihalo gas. This
angle increased as the protostellar evolution progressed, especially
as the star contracted to the main sequence, leading to a deeper
potential near the stellar surface and thus larger wind velocities.
The star formation efficiency due to protostellar outflow winds
remains near unity until $m_*\simeq 100\sm$, and then gradually
decreases to values of 0.3 to 0.7 by the time $m_*\simeq 300\sm$,
depending on the equatorial flattening of surrounding gas
distribution. Comparing these efficiencies to those from ionizing
feedback (\S\ref{S:feedback}), we conclude that the latter is more
important at determining the Pop III.1 IMF (see also Tan \& McKee 2008).

\section{How Accretion and Feedback Set the IMF}\label{S:feedback}

MT08 modeled the interaction of ionizing feedback on the accretion
flow to a Pop III.1 protostar. In the absence of WIMP annihilation
heating, the protostar contracts to the main sequence by the time
$m_*\simeq 100\sm$, and from there continues to accrete to higher
masses. At the same time, the ionizing luminosity increases, leading
to ionization of the infalling envelope above and below the plane of
the accretion disk. Once the \ion{H}{2} region has expanded beyond the
gravitational escape radius for ionized gas from the protostar,
pressure forces begin to act to reverse the infall. In the fiducial
case, by the stage when $m_*\simeq 100\sm$ we expect infall to have
been stopped from most directions in the minihalo. Only those regions
shadowed from direct ionizing flux from the protostar by the accretion
disk are expected to remain neutral and be able to accrete. 

In these circumstances the protostar starts to drive an ionized wind
from its disk (Hollenbach et al. 1994). Ionization
from the protostar creates an ionized atmosphere above the neutral
accretion disk, which then scatters some ionizing photons down on to
the shielded region of the outer disk, beyond $r_g$. An ionized
outflow is driven from these regions at a rate
\beq
\dot{m}_{\rm evap} 
\simeq 4.1\times 10^{-5} S_{\rm 49}^{1/2}T_{i,4}^{0.4}\msdt^{1/2}~~\smyr,
\eeq
where $S_{\rm 49}$ is the H-ionizing photon luminosity in units of
$10^{49}$ photons~$\rm s^{-1}$ and $T{i,_4}$ is the ionized gas temperature in units of $10^4$K. 

MT08 used the condition $\dot{m}_{\rm evap}>\dot{m}_*$ for determining
the final mass of the protostar. From numerical models they found it
is about $140 M_\odot$ in the fiducial case they considered, and
Table~\ref{tab:m} summarizes other cases. MT08 also made an analytic
estimate, assuming the H-ionizing photon luminosity is mostly due to the
main sequence luminosity of the star:
\beq
S\simeq 7.9\times 10^{49}\; \phi_S \mst^{1.5}~~~~{\rm ph\ s^{-1}},
\label{eq:ss}
\eeq which for $\phi_S=1$ is a fit to Schaerer's (2002) results,
accurate to within about 5\% for $60 M_\odot \lesssim m_*\lesssim 300
M_\odot$. Then the photoevaporation rate becomes 
\beq \dot{m}_{\rm evap} = 1.70\times 10^{-4}\phi_S^{1/2}(1+f_d)^{1/2} \left(\frac{T_{i,4}}{2.5}\right)^{0.4} \mst^{5/4}~\smyr.
\label{eq:evap2}
\eeq
The accretion rate onto the star-disk system is given by equation
(\ref{eq:mdot}). Equating this with equation (\ref{eq:evap2}), we find
that the resulting maximum stellar mass is
\beq
\label{eq:maxevap}
{\rm Max}\; m_{*f,2}=
6.3\;\frac{\esd^{28/47}\esdb^{12/47}
K'^{60/47}}{\phi_S^{14/47} (1+f_d)^{26/47}} 
        \left(\frac{2.5}{T_{i,4}}\right)^{0.24} \rightarrow 1.45,
\eeq
where the $\rightarrow$ assumes fiducial values $\esd=0.2$,
$\esdb=0.25$, $K'=1$, $\phi_S=1$, $f_d=1/3$, and $T_{i,4}=2.5$ (see MT08
for details; note also here in eq.~\ref{eq:maxevap} we have corrected
a sign error in the index for $\phi_S$). This analytic estimate
therefore also suggests that for the fiducial case ($K'=1$) the mass
of a Pop III.1 star should be $\simeq 140 M_\odot$.

The uncertainties in these mass estimates include: (1) the assumption
that the gas distribution far from the star is approximately spherical
--- in reality it is likely to be flattened towards the equatorial
plane, thus increasing the fraction of gas that is shadowed by the
disk and raising the final protostellar mass; (2) uncertainties in the
disk photoevaporation mass loss rate due to corrections to the
Hollenbach et al. (1994) rate from the flow starting inside $r_g$ and
from radiation pressure corrections; (3) uncertainties in the
\ion{H}{2} region breakout mass due to hydrodynamic instabilities and
3D geometry effects; (4) uncertainties in the accretion rate at late
times, where self-similarity may break down (Bromm \& Loeb 2004); (5)
the simplified condition, $\dot{m}_{\rm evap}>\mdsd$, used to mark the
end of accretion; (6) the possible effect of protostellar outflows
(discussed above); (7) the neglect of WIMP annihilation heating
(discussed above) and (8) the effect of rotation on protostellar
models, which will lead to cooler equatorial surface temperatures and
thus a reduced ionizing flux in the direction of the disk.

Here we discuss briefly the last of these effects. Using the results
of Ekstr\"om et al. (2008) and Georgi et al. (these proceedings), we
estimate that for a zero age main sequence protostar with
$\Omega/\Omega_{\rm crit}=0.99$ (i.e. rotating very close to
break-up), the surface temperature at an angle $80^\circ$ from the pole
(i.e. the direction relevant for the accretion disks modeled by MT08)
the surface temperature is reduced by a factor of 0.7. For
$m_*=140\sm$ this would cause $T_{\rm eff,*}$ to be reduced from
$1.0\times10^5\:{\rm K}$ to $7\times 10^4\:{\rm K}$ causing a
reduction in the ionizing flux (and thus also $\phi_S$) by a factor of
about 3. From eq.~\ref{eq:maxevap} we see that the mass of Pop III.1
star formation would be increased by about a factor of 1.4, to
$200\sm$ in the fiducial case.

\begin{table}
\begin{center}
\caption{Mass Scales of Population III.1 Protostellar Feedback}
\label{tab:m}
{\scriptsize
\begin{tabular}{cccccc}
\hline
$K^\prime$ & $f_{\rm Kep}$ & $T_{i,4}$ & $m_{\rm *,pb}$ ($\sm$)$^1$ & $m_{\rm *,eb}$ ($\sm$)$^2$ & $m_{\rm *,evap}$ ($\sm$)$^3$\\
\hline
1 & 0.5 & 2.5 & 45.3 & 50.4 & 137$^4$\\
\hline
1 & 0.75 & 2.5 & 37 & 41 & 137\\
1 & 0.25 & 2.5 & 68 & 81 & 143\\
1 & 0.125 & 2.5 & 106 & 170 & 173\\
1 & 0.0626 & 2.5 & 182 & 330$^5$ & 256\\
\hline
1 & 0.5 & 5.0 & 35 & 38 & 120\\
1 & 0.25 & 5.0 & 53.0 & 61 & 125\\
\hline
0.5 & 0.5 & 2.5 & 23.0 & 24.5 & 57\\
\hline
2.0 & 0.5 & 2.5 & 85 & 87 & 321\\
\hline
\end{tabular}
}
\end{center}
\vspace{1mm}
\scriptsize{
{\it Notes:}\\
$^1$Mass scale of HII region polar breakout.\\
$^2$Mass scale of HII region near-equatorial breakout.\\
$^3$Mass scale of disk photoevaporation limited accretion.\\
$^4$Fiducial model.\\
$^5$This mass is greater than $m_{\rm *,evap}$ in this case because it is calculated without allowing for a reduction in $\dot{m}_*$ during the evolution due to polar HII region breakout (see MT08).
}
\end{table}

\section{Evolution of the Pop III.1 IMF}


As the universe evolves and forms more and more structure, regions of
Pop III.1 star formation will become ever rarer. Indeed, because the
effects of radiation from previous stellar generations can propagate
relatively freely compared to the spreading and mixing of their metals
in supernovae, most metal-free star formation may be via Pop III.2
(Greif \& Bromm 2006). Nevertheless, understanding Pop III.1 star
formation is necessary as it establishes the initial conditions of
what follows.

O'Shea \& Norman (2007) studied the properties of Pop III.1
pre-stellar cores as a function of redshift. They found that cores at
higher redshift are hotter in their outer regions, have higher free
electron fractions and so form larger amounts of $\rm H_2$ (via $\rm
H^-$), although these are always small fractions of the total mass. As
the centers of the cores contract above the critical density of
$10^4\:{\rm cm^{-3}}$, those with higher $\rm H_2$ fractions are able
to cool more effectively and thus maintain lower temperatures to the
point of protostar formation. The protostar thus accretes from
lower-temperature gas and the accretion rates, proportional to $c_s^3
\propto T^{3/2}$, are smaller. Measuring infall rates at the time of
protostar formation at the scale of $M=100\sm$, O'Shea \& Norman find
accretion rates of $\sim 10^{-4}\smyr$ at $z=30$, rising to $\sim
2\times 10^{-2}\smyr$ at $z=20$. If Hunter's (1977) solution applies,
the mass accretion rates to the protostar will be higher by a factor
of 3.7 by the time $\msd=100\sm$. These accretion rates then correspond to $K'$=0.37 ($z=30$) to
4.3 ($z=20$). A naive application of eq.~\ref{eq:maxevap} would imply
a range of masses of $40\sm$ to $900\sm$. This suggests that the very
first Pop III.1 stars were relatively low-mass massive stars,
e.g. below the mass required for pair instability supernovae
($140-260\;M_\odot$ in the models of Heger \& Woosley 2002). Such
stars would have had relatively little influence on their cosmological
surroundings, thus allowing Pop III.1 star formation to continue to
lower redshifts. It is not yet clear from simulations when Pop III.1
star formation was finally replaced by other types, since this depends
on the early IMFs of Pop III.1, III.2 and II stars. This transition
presumably occurred before reionization was complete.

\acknowledgements
We thank the organizers of IAU255 for a very stimulating meeting. The research of JCT is supported by NSF CAREER grant AST-0645412.

\end{document}